\documentclass[aps,showpacs,twocolumn,preprintnumbers,amssymb,amsmath,amsfonts,superscriptaddress]{revtex4}
\pdfoutput=1
\usepackage{graphicx}
\usepackage{subfigure}
\usepackage{color}

\begin{document}

\title{Scanning Tunneling Spectroscopy and Vortex Imaging\\ in the Iron-Pnictide Superconductor BaFe$_{1.8}$Co$_{0.2}$As$_2$}

\author{Yi Yin}
\affiliation{Department of Physics, Harvard University, Cambridge,
MA 02138, U.~S.~A.}
\author{M. Zech}
\affiliation{Department of Physics, Harvard University, Cambridge,
MA 02138, U.~S.~A.}
\author{T. L. Williams}
\affiliation{Department of Physics, Harvard University, Cambridge,
MA 02138, U.~S.~A.}
\author{X. F. Wang}
\affiliation{Hefei National Laboratory for Physical Science at
Microscale and Department of Physics, University of Science and
Technology of China, Hefei, Anhui 230026, P.~R.~China}
\author{G. Wu}
\affiliation{Hefei National Laboratory for Physical Science at
Microscale and Department of Physics, University of Science and
Technology of China, Hefei, Anhui 230026, P.~R.~China}
\author{X. H. Chen}
\affiliation{Hefei National Laboratory for Physical Science at
Microscale and Department of Physics, University of Science and
Technology of China, Hefei, Anhui 230026, P.~R.~China}
\author{J. E. Hoffman}
 \email{jhoffman@physics.harvard.edu}   
\affiliation{Department of Physics, Harvard University, Cambridge,
MA 02138, U.~S.~A.}
\date{\today}

\begin{abstract}
We present an atomic resolution scanning tunneling spectroscopy study of
superconducting BaFe$_{1.8}$Co$_{0.2}$As$_2$ single crystals in
magnetic fields up to $9\,\text{Tesla}$. At zero field, a single gap with coherence peaks at $\overline{\Delta}=6.25\,\text{meV}$ is
observed in the density of states.  At $9\,\text{T}$ and $6\,\text{T}$, we image a disordered vortex lattice, consistent with isotropic, single flux quantum vortices. Vortex locations are uncorrelated with strong scattering surface impurities, demonstrating bulk pinning.  The vortex-induced sub-gap density of states fits an exponential decay from the vortex center, from which we extract a coherence length $\xi=27.6\pm 2.9\,\text{\AA}$, corresponding to an upper critical field $H_{c2}=43\,\text{T}$.
\end{abstract}


\maketitle

The recently discovered iron-arsenic based superconductors (pnictides) break the
monopoly that copper oxides (cuprates) have held over high temperature superconductivity for more than two decades~\cite{KamiharaJAMCS2008, XHChenNature2008, GFChenPRL2008, ZARenEPL2008, RotterPRL2008}. There is new hope that investigation of the pnictides as a `foil' for cuprates will finally lead to a fundamental understanding of high-$T_c$ superconductivity.  An immediate challenge is therefore to catalogue and understand the similarities and differences between cuprates and pnictides.  Both materials become superconducting after chemical doping of anti-ferromagnetic parent compounds and both have dome-shaped phase
diagrams. Spin-fluctuation-mediated pairing, a promising mechanism for cuprate superconductivity, has also been proposed for the pnictides~\cite{MazinPRL2008, Wang}. However, significant differences are thought to exist between these two families of materials. For example, the parent compounds of pnictide superconductors are computed to be semi-metals~\cite{SinghPRL2008} instead of Mott insulators.  Of more technological relevance, the pnictides are found to be more isotropic in a magnetic field~\cite{NiPRB2008}, which may facilitate application due to more effective pinning of quantized magnetic vortices~\cite{Physics.1.21}.  A second challenge is therefore to experimentally characterize vortex pinning.

To investigate the pnictides, both in comparison to the cuprates, and for applications requiring vortex pinning, we use a home-built cryogenic scanning tunneling microscope (STM) capable of tracking atomically resolved locations as the magnetic field is swept up to $9\,\text{T}$. We choose to study optimally doped single crystal BaFe$_{1.8}$Co$_{0.2}$As$_2$. Since the Co dopants are incorporated into the strongly bound FeAs layer, the topmost layer is more like the bulk and remains more stable while tunneling. Our samples, grown with FeAs flux to avoid contamination by other ele\-ments~\cite{Wang:arXiv0806.2452}, show a sharp resistive transition at $T_c=25.3\,\text{K}$ with width $\Delta T=0.5\,\text{K}$.

Figure 1 shows an atomic resolution topographic image of the surface of BaFe$_{1.8}$Co$_{0.2}$As$_2$, cleaved \textit{in situ} at $\sim$~$\!25\,\text{K}$ and recorded at $6.15\,\text{K}$ in zero magnetic field. From the Fourier transformation we find an interatomic spacing consistent with x-ray diffraction measurements
\begin{figure}[b]
\begin{center}
\includegraphics[width=0.85\columnwidth]{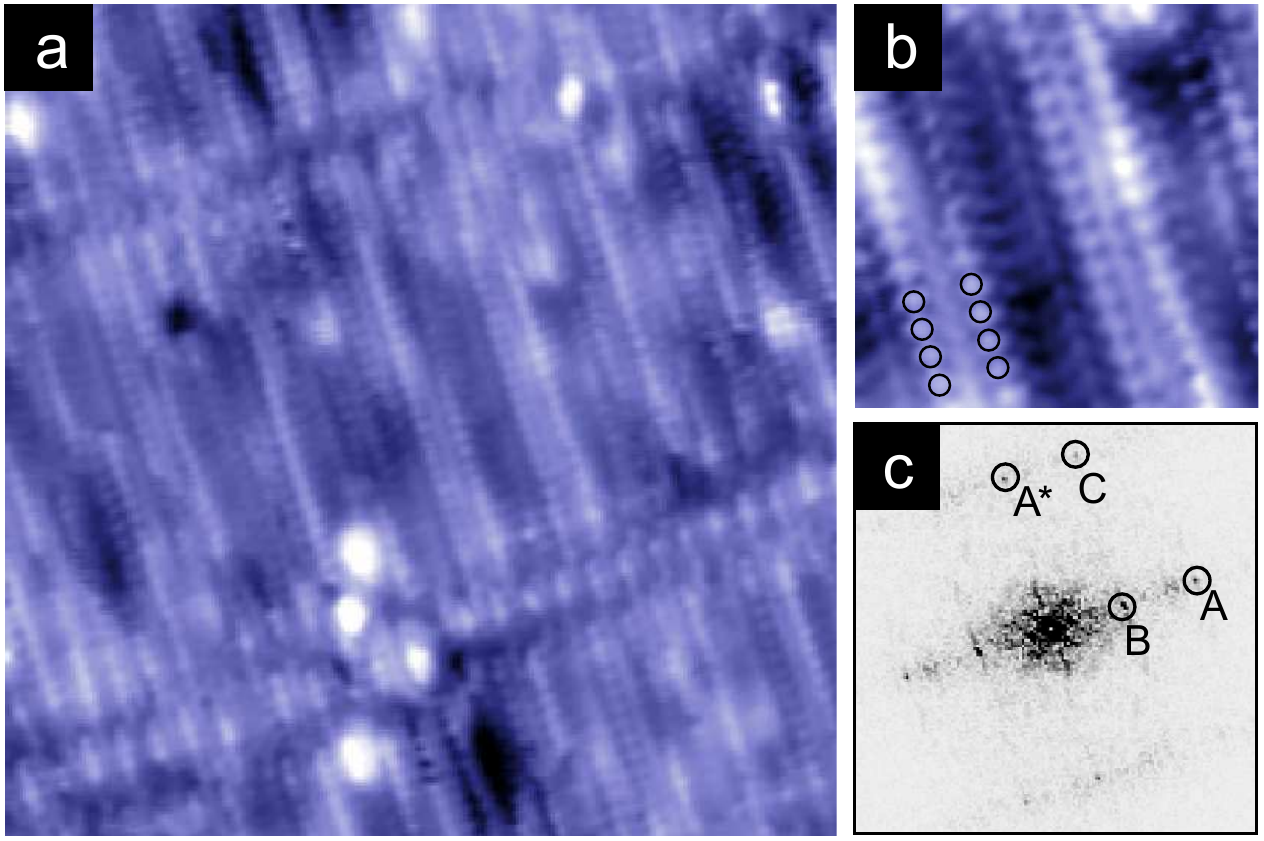}
\caption{(color online). (a) A $20\times20\,\text{nm}^2$ constant-current topographic image of the cleaved surface of BaFe$_{1.8}$Co$_{0.2}$As$_2$. The interatomic spacing is $a=3.96\,\text{\AA}$. (b) Zoom in ($\times 2$ magnification) on a $5\times5\,\text{nm}^2$ area within (a).
The black circles denote the positions of the $2\times 1$ stripe structure. (c) Fourier transformation of (a).  Peaks $A$ and $A^\star$ represent the atomic spacing between and within rows, respectively. Peaks $B$ and $C$ represent the alternating row characteristics: $B$ shows intensity modulations attributable to height or density of states variations, while $C$ shows in-plane shifts along the row direction.
Both images are recorded using $V_{\mathrm{sample}}=-20\,\text{mV}$ and $I=40\,\text{pA}$, at $T=6.15\,\text{K}$ in zero magnetic field.}
\end{center}
\end{figure}\noindent in either the Ba or As layer, $a=3.96\,\text{\AA}$~\cite{SefatPRL2008}. The surface shows a complex stripe-like structure in which alternate rows appear shifted with respect to each other. Similar stripes were observed in a previous STM study on Sr$_{1-x}$K$_x$Fe$_2$As$_2$~\cite{Boyer:arXiv0806.4400}. The stripe-like feature is oriented at $45^\circ$ to the one-dimensional spin order observed in the parent compound of these materials~\cite{delaCruz} and is therefore not likely related to this bulk order.
In addition to the stripe-like feature, there is another sparse, perpendicular, one-dimensional feature in the topographic image which is a possible dislocation to relieve surface strain. Neither feature has shown any effect on the local superconducting gap.

We focus first on electron pairing, the fundamental origin of superconductivity, which is characterized by the superconducting energy gap $\Delta$ in the electronic density of states (DOS). We record differential conductance $dI/dV$ as a function of sample-tip voltage $V$, which is proportional to the local density of states $g(\vec{r},eV)$. For each $dI/dV$ spectrum in a dense array of locations, we extract the magnitude of the energy gap $\Delta$, one half the distance between coherence peaks. A single energy gap $\Delta\sim6\,\text{meV}$ is observed in all spectra, independent of junction resistance. This gap energy is comparable to the nodeless gap on the $\Gamma$-centered Fermi surface ($\beta$ FS) reported by angle resolved photoemission spectroscopy (ARPES) on similar electron doped BaFe$_{1.85}$Co$_{0.15}$As$_2$~\cite{Ding2008}.

\begin{figure}[b]
\begin{center}
\includegraphics[width=0.85\columnwidth]{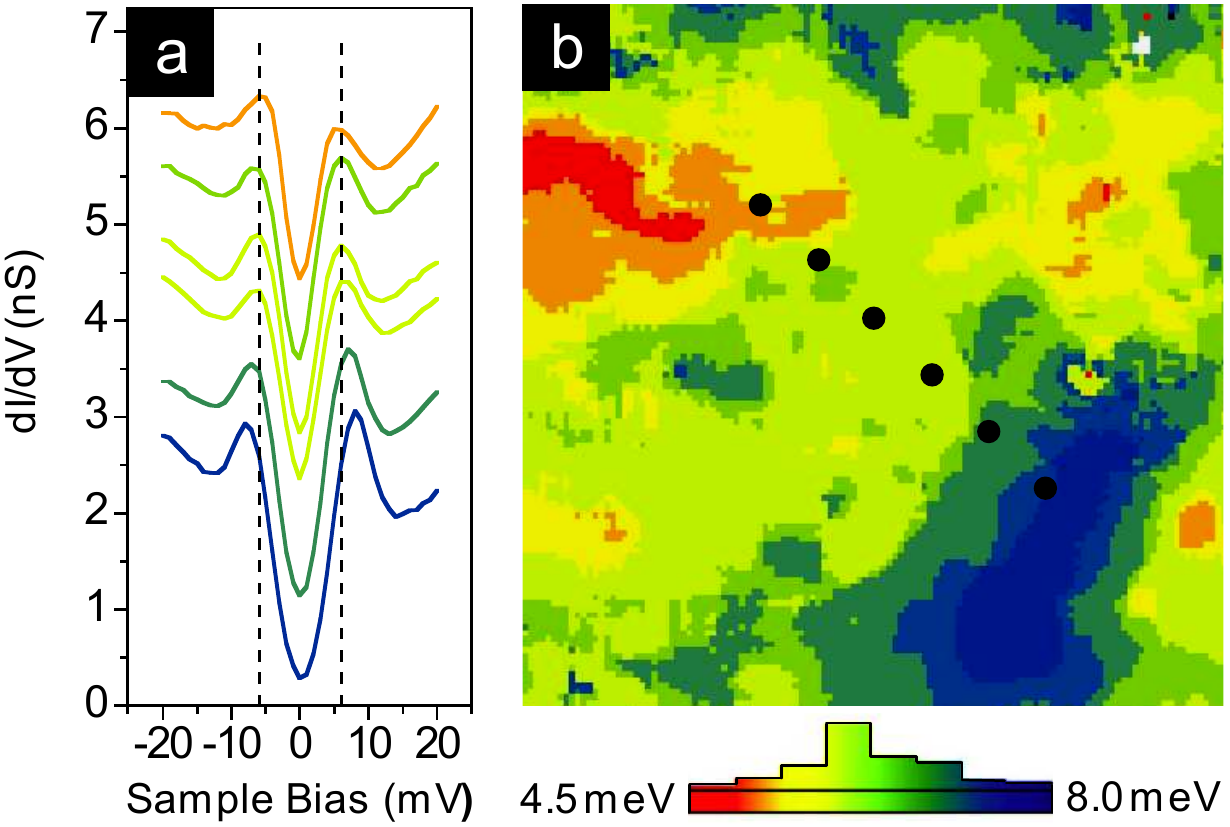}
\caption{(color online). (a) A series of $dI/dV$ spectra taken along an $11\,\text{nm}$ line, illustrating the shape of the superconducting gap and the low differential conductance at the Fermi energy. To reduce noise, each spectrum shown is the average of all spectra acquired within a $5\,\text{\AA}$ radius. The vertical dashed lines are guides to the eye located at $\pm6\,\text{meV}$. (b) A $20\times20\,\text{nm}^2$ gap map, revealing the spatial variation of  the gap magnitude $\Delta$. The gap has $\overline{\Delta}=6.25\,\text{meV}$ and fractional variation only $12\,\%$. A color-coded histogram of $\Delta$ is shown below the gap map. The six spectra in (a) (from top to bottom) are taken at the locations of the black points indicated in (b) (from upper left to lower right). Data were acquired at $T=6.25\,\text{K}$ in zero magnetic field.}
\end{center}
\end{figure}

A map of the spatial dependence of $\Delta$ (Fig.\ 2b) at zero magnetic field and $6.25\,\text{K}$ shows a total range of $\Delta$ from $4.5$ to $8.0\,\text{meV}$. From the $\sim$16,000 measured spectra, we extract the average $\overline{\Delta}=6.25\,\text{meV}$ and standard deviation $\sigma=0.73\,\text{meV}$, leading to a fractional variation $\sigma/\overline{\Delta}$ of $12\,\%$. Spectra with smaller gap energies tend to exhibit weaker coherence peaks and higher zero bias conductance, as exemplified in Fig.\ 2a. This effect may be explained by impurity scattering in unconventional superconductors~\cite{Muzikar1994}. In Bi$_2$Sr$_2$CaCu$_2$O$_{8+x}$ (Bi2212), gaps with  $\overline{\Delta}\sim33\,\text{meV}$ and $\sigma\sim7\,\text{meV}$ ($\sigma/\overline{\Delta}\sim 21\,\%$) are typically reported~\cite{mcelroy:197005}. In Bi-based cuprates, the larger intrinsic electronic inhomogeneity and the observed (opposite) relation between coherence peak height and gap width may be complicated by the coexistence of a pseudogap at nearby energy~\cite{Hudson, Madhavan2008}.

\begin{figure}[b]
\begin{center}
\includegraphics[width=0.95\columnwidth]{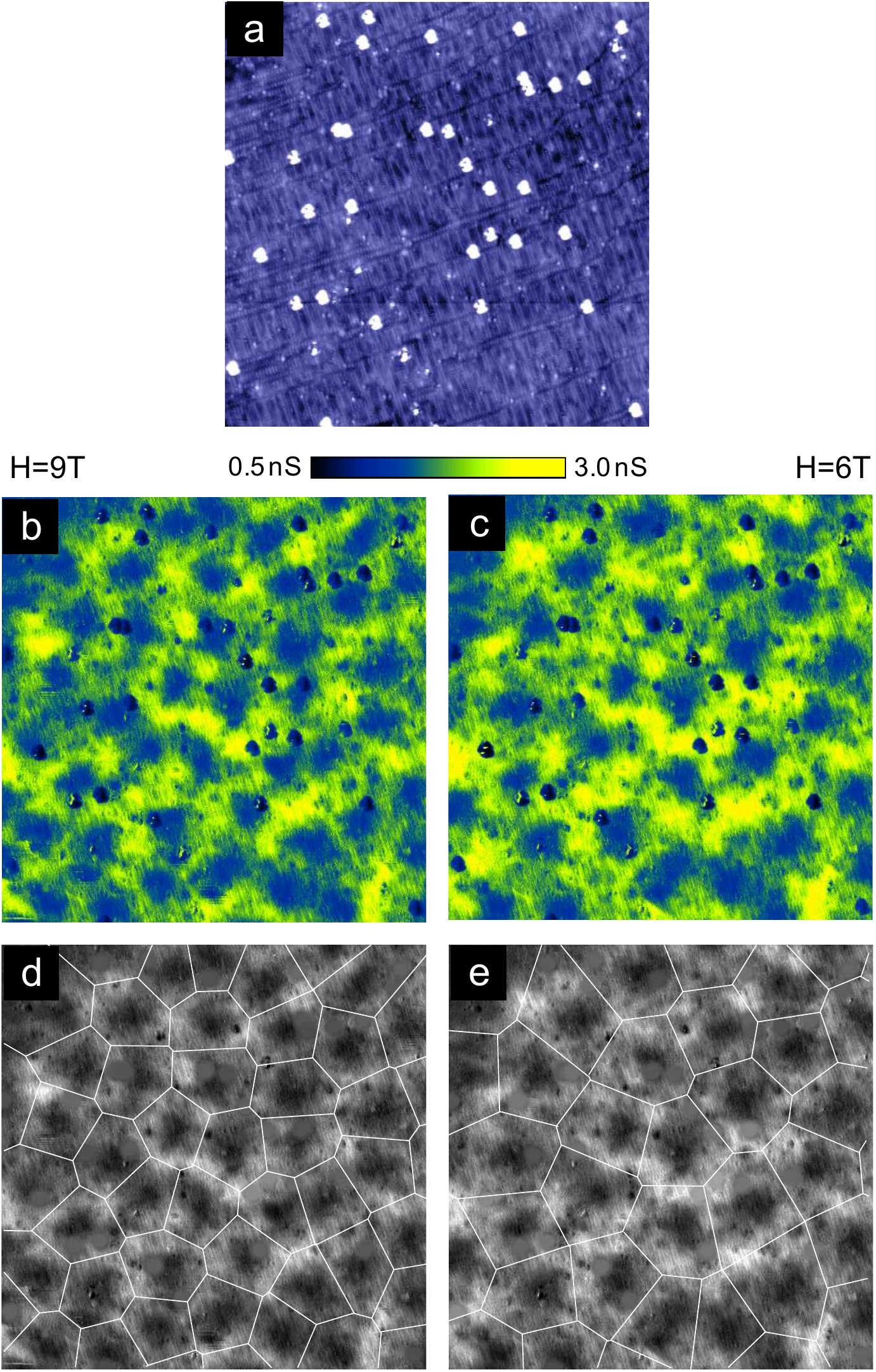}
\caption{(color online). (a) A $100\times100\,\text{nm}^2$ topographic image recorded in $9\,\text{T}$ magnetic field with $V_{\mathrm{sample}}=-5\,\text{mV}$ and $I=10\,\text{pA}$. The bright spots are impurities. (b) Differential conductance map recorded simultaneously with the topographic image in (a), revealing the sample DOS at $5\,\text{meV}$ in $9\,\text{T}$ magnetic field at $T=6.15\,\text{K}$. Vortices are visible as dark features due to the local suppression of the coherence peaks. The vortices form a disordered lattice.  Impurities, also visible in (a), appear as smaller, sharper dark features. (c) Differential conductance map recorded at $6\,\text{T}$ in the same field of view and with the same parameters as (b), except $T=6.21\,\text{K}$. (d, e) A Voronoi overlay on impurity-filtered data emphasizes the vortices and demonstrates the procedure used to compute the average area associated with each vortex.}
\end{center}
\end{figure}

From the average of the gap map we calculate the reduced gap $2\overline{\Delta}/k_\text{B}T_c=5.73$. This exceeds the values for weak-coupling $s$-wave or $d$-wave BCS superconductors, which are 3.5 and 4.3, respectively~\cite{Tesanovic, PhysRevB.49.1397}. Experiments which decouple the pseudogap by coherent tunneling~\cite{AndreevRP} or normalization~\cite{Hudson} find reduced gap values in the cuprates between 6 and 10. Our result suggests that although pnictides are in a strong coupling regime, they are not as strongly coupled as cuprates.

While bulk studies have indicated high critical fields in the pnictides~\cite{Yamamoto0810.0699}, single vortex imaging gives the most stringent bounds on vortex pinning~\cite{StraverAPL}. Static vortices have been imaged magnetically by scanning SQUID microscopy at $33\,\text{mG}$ in the related compound NdFeAsO$_{0.94}$F$_{0.06}$~\cite{Hicks:arXiv0807.0467}, but it remains important to understand pinning at the higher fields which will be of interest for applications such as superconducting magnets.

We image vortices electronically by mapping the conductance at an energy where a vortex alters the density of states. Fig.~3b shows such a conductance map, recorded in a $9\,\text{T}$ magnetic field at an energy corresponding to the filled state coherence peak. A second conductance map at the same location, recorded at $6\,\text{T}$, is shown in Fig.~3c. In both maps, the vortices appear as broad areas of depressed conductance. Impurities, possibly single Fe or Co vacancies, appear as sharper minima in the conductance, also visible as white spots in the simultaneously recorded topography in Fig.~3a (at $9\,\text{T})$. As in YBa$_2$Cu$_3$O$_{7-x}$ (Y123) and Bi2212, the vortex lattice is disordered~\cite{PhysRevLett.75.2754, PanPRL2000}. We find no correlation between the locations of the vortices and the visible impurities~\cite{Impurities}; therefore bulk pinning must play a dominant role in this material. These observations are not compatible with pancake vortices~\cite{GrigorenkoNature}, which would be pinned by the surface impurities. The isotropy of the Fourier transformation of the vortex lattice also indicates that vortex locations are not significantly affected by the stripe-like surface feature, or by any residual one-dimensional spin or structural order~\cite{delaCruz} within the bulk.

After filtering out the impurities, a peak-fitting algorithm was used to extract the vortex center locations from the conductance maps. Voronoi cells~\cite{PanPRL2000} were overlaid onto the vortex image (Fig.~3, d and e) and the average per-vortex flux $\overline{\phi}$ was calculated from the Voronoi cell size. We observe $\overline{\phi}(9\mathrm{T})=2.05\times 10^{-15}\, \mathrm{T}\,\mathrm{m}^2$ and $\overline{\phi}(6\mathrm{T})=2.17\times 10^{-15}\, \mathrm{T}\,\mathrm{m}^2$, in good agreement with the single magnetic flux quantum, $\Phi_0=2.07\times 10^{-15}\, \text{T}\,\text{m}^2$.

\begin{figure}[b]
\begin{center}
\includegraphics[width=1\columnwidth]{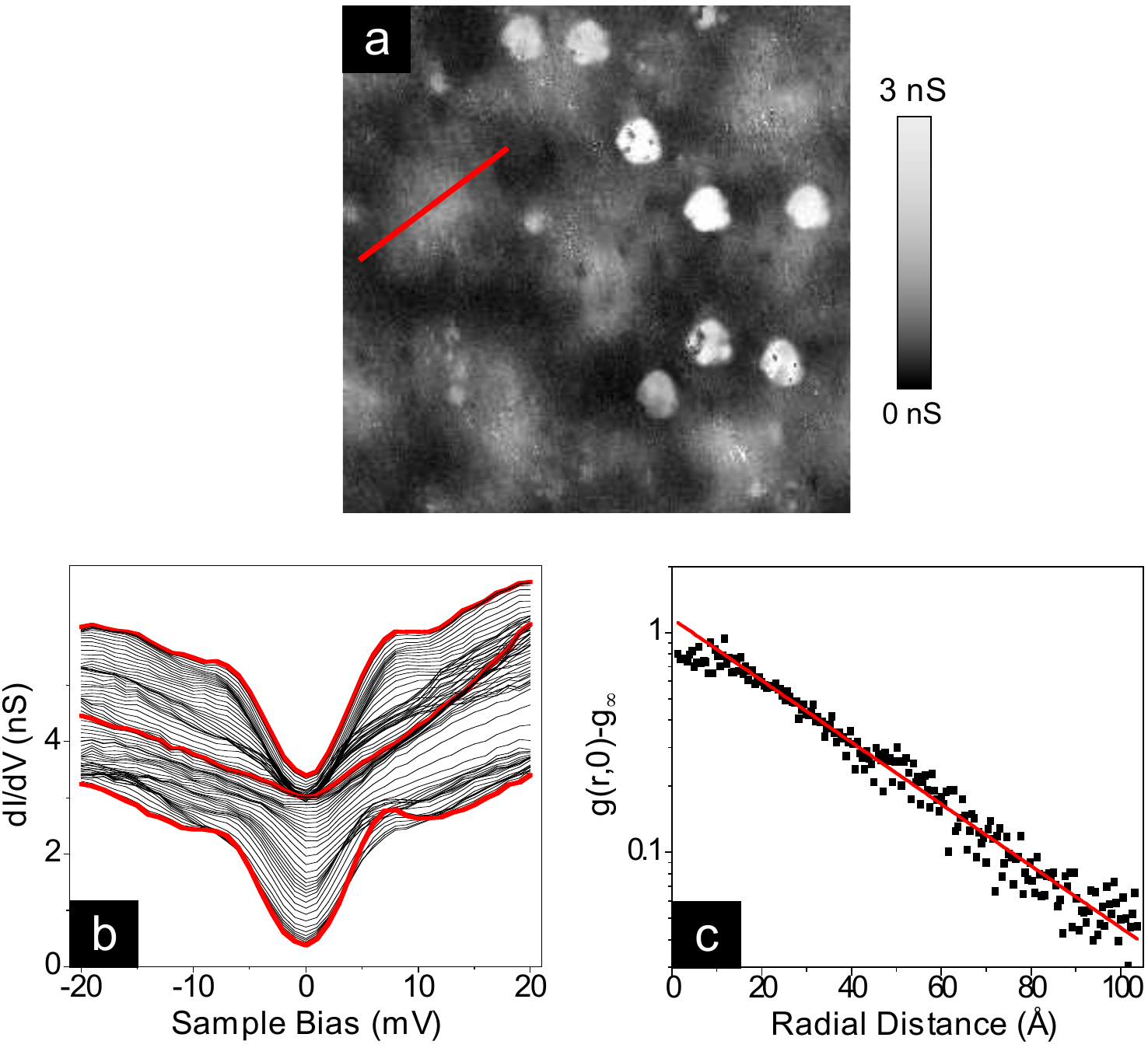}
\caption{(color online). (a) A $40\times40\,\text{nm}^2$ map of the zero bias conductance in $9\,\text{T}$ magnetic field at $T=6.15\,\text{K}$, showing $\sim\!\!8$ vortices (broader, lighter objects) and 8 strong-scattering impurities (sharper, brighter objects). (b) A series of spectra along a $14.5\,\text{nm}$ trajectory through one of the vortices, indicated by the red line in (a). The superconducting gap is completely extinct inside the vortex core; only the $\bigvee$-shaped background remains. The three thick red lines emphasize the spectra at the vortex core and far from it on both sides. The 75 spectra shown are offset by a total of $3\,\text{nS}$ for clarity. (c) Azimuthally averaged radial dependence of the differential conductance $g(r,0)$ around a single vortex as measured (black squares) and fit by an exponential decay (red line). The constant background $g_\infty$ has been removed in order to emphasize the exponential decay on the logarithmic scale of the $y$-axis.  The exponential fit leads to a coherence length of $\xi =30.8\pm1.4\,\text{\AA}$ for this vortex and to an average of $\xi=27.6\,\text{\AA}$ with standard deviation $2.9\,\text{\AA}$ for all vortices investigated.}
\end{center}
\end{figure}

We gain insight into a phenomenon by probing the conditions under which it breaks down: local measurements in high magnetic field allow us to observe the destruction of superconductivity in vortex cores. Fig.~4a shows the zero bias conductance (ZBC), corresponding to the Fermi level density of states, in a region of Fig.~3b. Here the vortices appear as enhanced sub-gap density of states, in contrast to Figs.~3b-e, where the vortices appear as depressed conductance due to the local suppression of the coherence peaks. Fig.~4b shows a series of spectra on a trajectory through one of the vortices in $9\,\text{T}$ magnetic field. No ZBC peaks, which would be a signature of quasiparticle bound states at energies $\frac{1}{2}\Delta^2/{E_F}$~\cite{Caroli1964}, are observed within the vortex core as would be expected in a conventional $s$-wave superconductor; only the $\bigvee$-shaped background remains. In Y123~\cite{PhysRevLett.75.2754} and Bi2212~\cite{PanPRL2000}, particle-hole symmetric subgap peaks have been reported within vortex cores, with energies approximately $\pm\Delta/4$.  We do not observe such particle-hole symmetric subgap states, although the corresponding low energy scale in this material, $\Delta/4\sim1.5\,\text{meV}$ would likely appear as a weak ZBC peak as well. Our observed pnictide vortices are isotropic and lacking internal structure, in contrast to predictions of $d$-wave vortices~\cite{Franz} and observations of 4-fold symmetric internal vortex structure in $d$-wave cuprates~\cite{Hoffman}.

To extract the superconducting coherence length $\xi$, we use the azimuthally-averaged radial dependence of the vortex-induced ZBC.  We fit an exponential decay $g(r,0)= g_\infty+A\exp(-r/{\xi})$ within a distance of $r=20$ to $100\,\text{\AA}$ from the vortex core, as shown in Fig. 4c. From several vortices, we find an average coherence length $\xi=27.6\,\text{\AA}$ with standard deviation $2.9\,\text{\AA}$. Using the Ginzburg-Landau expression $H_{\text{c2}}=\phi_0/2\pi\xi^2$, we compute the upper critical field $H_{c2}=43\,\text{T}$.

With residual resistivity $\rho_0=0.23\,\text{m}\Omega\,\text{cm}$ and Hall coefficient $R_{H}=11\times10^{-9}\,\text{m}^3/\text{C}$, the electronic mean free path $\ell=\hbar(3\pi^2)^{1/3}/e^2n^{2/3}\rho_0$ is approximately $81\,\text{\AA}$. Given that our coherence length  $\xi=27.6\,\text{\AA}$ is almost three times smaller than $\ell$, the superconductor under investigation is not in the dirty regime in which the ZBC peak would be suppressed due to significant scattering.

In contrast to the cuprates, where substitutions into the critical superconducting CuO$_2$ plane lead to strong scattering and the reduction of $T_c$~\cite{HudsonNickel,TarasconPRB1987}, the 10\% of Co atoms doped directly into the superconducting FeAs plane in this material do not provide a strong scattering signature.  Strong scatterers are dilute in this material, corresponding to only 0.029\% of the Fe sites, leading to an average spacing comparable to $\ell\sim81\,\text{\AA}$.

To summarize, we have presented the first atomically resolved STM/STS studies on a single crystal pnictide superconductor in magnetic fields up to $9\,\text{T}$.  We observe a single gap with $\overline{\Delta}=6.25\,\text{meV}$ and relative standard deviation $12\,\%$ in zero magnetic field.  We have imaged a stationary but disordered vortex lattice at $6\,\text{T}$ and $9\,\text{T}$, uncorrelated with the locations of surface impurities. This demonstrates that vortices experience bulk pinning at fields up to $9\,\text{T}$, even in a clean single crystal pnictide superconductor with $T_c$ only $25.3\,\text{K}$.  The non-observation of sub-gap peaks within the vortex cores contrasts with observations in both $d$-wave cuprates and conventional $s$-wave superconductors.

We acknowledge Eric Hudson, Subir Sachdev, and Ophir Auslaender for valuable discussions. This work is supported by the NSF (DMR-0508812) and the AFOSR (FA9550-05-1-0371).  T.L.W. acknowledges support from an NDSEG fellowship.


\begin{thebibliography}{34}
\bibitem[{Imp()}]{Impurities}
\bibinfo{note}{At $9\,\text{T}$, 10.6\% of the total field of view is covered
  by vortex cores, and 4 out of 49 impurities lie within these cores. The
  difference between 10.6\% and 4 out of 49 (8.2\%) is not statistically
  significant.}
\end{thebibliography}

\end{document}